\documentclass[twocolumn,showpacs,floatfix,aps,pra]{revtex4}
\usepackage{graphicx}
\usepackage{dcolumn}
\usepackage{bm}

\begin{document}
\title{Multiphoton antiresonance}

\author{ M. I. Dykman$^{1}$ and M. V. Fistul$^{2}$}
 \affiliation{
$^{1}$ Department of Physics and Astronomy, Michigan State
 University, East Lansing, MI 48824, USA\\
$^{2}$ Theoretische Physik III,
Ruhr-Universit\"at Bochum, D-44801 Bochum, Germany }
\date{\today}

\begin{abstract}
We show that nonlinear response of a quantum oscillator displays
antiresonant dips and resonant peaks with varying frequency of the
driving field.  The effect is a consequence of special symmetry and is
related to resonant multiphoton mixing of several pairs of oscillator
states at a time. We discuss the possibility to observe the
antiresonance and the associated multiphoton Rabi oscillations in
Josephson junctions.
\end{abstract}


\maketitle

Many resonant nonlinear phenomena are described by the model of a
nonlinear oscillator in a resonant field. Examples include
collisionless dissociation of molecules \cite{Bloembergen76},
dispersive optical bistability \cite{Drummonds80}, cyclotron resonance
of a relativistic electron \cite{Gabrielse85}, resonant enhancement of
activated escape \cite{Larkin_all} and tunneling \cite{JosEscape} in
Josephson junctions, and recently discovered hysteresis in Josephson
junctions \cite{Siddiqi_ampl03} and nanomechanical resonators
\cite{Cleland04}.

A weakly nonlinear oscillator is a multi-level quantum system with
nearly equidistant energy levels $E_n$. Therefore a periodic force of
frequency $\omega_F$ can be nearly resonant for many transitions at a
time, i.e., $\hbar\omega_F$ can be close to the interlevel distance
$E_{n+1}-E_n$ for many $n$. This makes an oscillator convenient for
studying multiphoton Rabi oscillations. They arise when the spacing
between remote energy levels $n$ and $m$ coincides with the energy of
$n-m$ photons, $E_n-E_{m} = (n-m)\hbar\omega_F$
\cite{Bloembergen76}. The multiphoton transition amplitude is
resonantly enhanced, because the $m\to n$ transition occurs via a
sequence of virtual field-induced transitions $k\to k+1$ (with $m\leq
k \leq n-1$), all of which are almost resonant.

In this paper we show that multiphoton transitions in the oscillator
are accompanied by a new effect, an {\it antiresonance} of the
response.  When the frequency of the driving field adiabatically
passes through a resonant value, the vibration amplitude displays a
sharp minimum or maximum, depending on the initial conditions. We
argue that the antiresonance and the multiphoton Rabi oscillations can
be observed in such macroscopic systems as Josephson junctions and
nanomechanical resonators.

The multiphoton antiresonance is a consequence of two interesting
properties of a driven oscillator. First, for resonant photon
frequencies, simultaneously, not one but several pairs of states turn
out to be in resonance. Second, the amplitudes of forced vibrations in
the resonating states coincide with each other, in the neglect of
multiphoton mixing. For not too strong fields, these amplitudes are
determined by nonresonant field-induced coupling of neighboring
Fock states of the oscillator. The resonant multiphoton mixing leads
to level splitting, which strongly differs for different resonating
pairs. It is this difference that results in the dips (peaks) in the
vibration amplitude as $\omega_F$ adiabatically passes through
resonances.

In the semiclassical picture, resonant multiphoton transitions
correspond to tunneling between Floquet states of the oscillator with
equal quasienergies [the quasienergy $\varepsilon$ gives the change of
the wave function $\psi(t)$ when time is incremented by the modulation
period $\tau_F$,
$\psi(t+\tau_F)=\exp(-i\varepsilon\tau_F/\hbar)\psi(t)$]. The
occurrence of equal-quasienergy states is related to the bistability
of forced vibrations of a classical oscillator
\cite{LL-Mechanics}. There is similarity between the oscillator
tunneling and the tunneling of a particle in a static double-well
potential, with the potential minima being analogs of the stable
states of forced vibrations. However, the latter states are not
separated by a static barrier.  Tunneling of a driven oscillator
\cite{Dyakonov86} is a carefully studied example of dynamical
tunneling \cite{Heller81}. As we show, the WKB analysis gives an
important insight into the origin of the antiresonance.

The Hamiltonian of a driven nonlinear oscillator with mass $M=1$ has
the form
\begin{equation}
\label{Hamiltonian_full}
H(t)={1\over 2}p^2 + {1\over 2}\omega_0^2q^2 + {1\over 4}\gamma q^4
-qA\cos(\omega_Ft).
\end{equation}
We assume that the driving field is nearly resonant, i.e., the
frequency detuning $\delta\omega$ is small,
\begin{equation}
\label{detuning}
|\delta\omega| \ll \omega_F, \qquad \delta\omega = \omega_F-\omega_0.
\end{equation}

We consider not too large amplitudes of the driving field $A$, so that
the oscillator anharmonicity is small, and in particular $|\gamma|
q^2\ll \omega_0^2$ for typical $q$.  We also assume that $\gamma$ and
$\delta\omega$ have the same sign. This is required for the amplitude
of classical forced vibrations to display hysteresis as a function of
the force amplitude $A$. If there is a cubic term $\alpha q^3/3$ in
the potential energy of the oscillator, its major effect of interest
for this paper is the renormalization $\gamma \to \gamma -\frac{10}{9}
(\alpha/\omega_0)^2$ \cite{LL-Mechanics}.

To study quantum dynamics, we will write the Hamiltonian in terms
of the raising and lowering operators of the oscillator $a^{\dagger},
a$, and switch to the rotating frame with a canonical transformation
$U(t)=\exp(-i\omega_F\,a^{\dagger}a\, t)$. The transformed Hamiltonian
$H_0=U^{\dagger}(t)H(t)U(t) -
i\hbar U^{\dagger}(t)\dot U(t)$
is time-independent in the rotating wave approximation (RWA),
\begin{eqnarray}
\label{Hamiltonian_working}
H_0=-\delta\omega \hat n \! &+& \! {1\over
2}V \hat n(\hat n + 1) -
f\left(a+a^{\dagger}\right), \quad \hat n=
a^{\dagger}a,
\nonumber\\
V&=&3\hbar\gamma/4\omega_0^{2},
\qquad f=(8\hbar\omega_0)^{-1/2}A.
\end{eqnarray}
The terms $\propto V,f$ that contain fast-oscillating factors $\propto
\exp(\pm ik\omega_Ft)$ with $k=2,4$ were disregarded in
Eq.~(\ref{Hamiltonian_working}). In the expression for $H_0$ and in
what follows $\hbar =1$.

The eigenvalues of the Hamiltonian (\ref{Hamiltonian_working})
$\varepsilon_n$ give the quasienergies of the driven oscillator.
In the limit of weak driving their spectrum is particularly simple,
\begin{equation}
\label{fto0}
\varepsilon_n=-n\delta\omega + Vn(n+1)/2 \qquad (f\to
0).
\end{equation}

We will study multiphoton resonance for the ground state of the
oscillator, $E_N-E_0=N\omega_F$, or equivalently
$\varepsilon_0=\varepsilon_N$. From Eq.~(\ref{fto0}), for small $f$
and given $N$ the resonance occurs for
\[\delta \omega=\delta\omega_N\equiv V(N+1)/2.\]
Remarkably, at resonance {\it all} quasienergy levels (\ref{fto0})
with $n\leq N$ are pairwise-degenerate,
$\varepsilon_{N-n}=\varepsilon_n$ \cite{Risken-Walls87}. Equivalently,
$E_{N-n}-E_n=(N-2n)\omega_F$. One can see that the degeneracy is not
lifted by the lowest-order ($\propto f^2$) field-induced level shift
[except for the levels $n=(N\pm 1)/2$ for odd $N$ and $n=(N/2)\pm 1$
for even $N$]. As shown below, it persists for all $f$ in the WKB
approximation in the neglect of tunneling.

The response of the oscillator to the field is characterized by the
expectation value of its coordinate $q$. If the oscillator is in an
eigenstate $|n\rangle$ of the Hamiltonian (\ref{Hamiltonian_working}),
\begin{equation}
\label{a_n}
q_n=(2\omega_0)^{-1/2}a_ne^{-i\omega_Ft}+{\rm c.c.},
\qquad a_n=\langle n|a|n\rangle.
\end{equation}

To first order in the field, the reduced amplitude of forced
vibrations $a_n$ is
\begin{equation}
\label{a_first_order}
a_n=-f\delta\omega/ \bigl((\delta\omega -Vn)[\delta\omega-V(n+1)]\bigr).
\end{equation}
At multiphoton resonance, where $\delta\omega =\delta\omega_N$, the
vibration amplitudes in the resonating states coincide with each
other, $a_{N-n}=a_n$ for $0\leq n <N/2$.

\begin{figure}[h]
\includegraphics[width=2.5in]{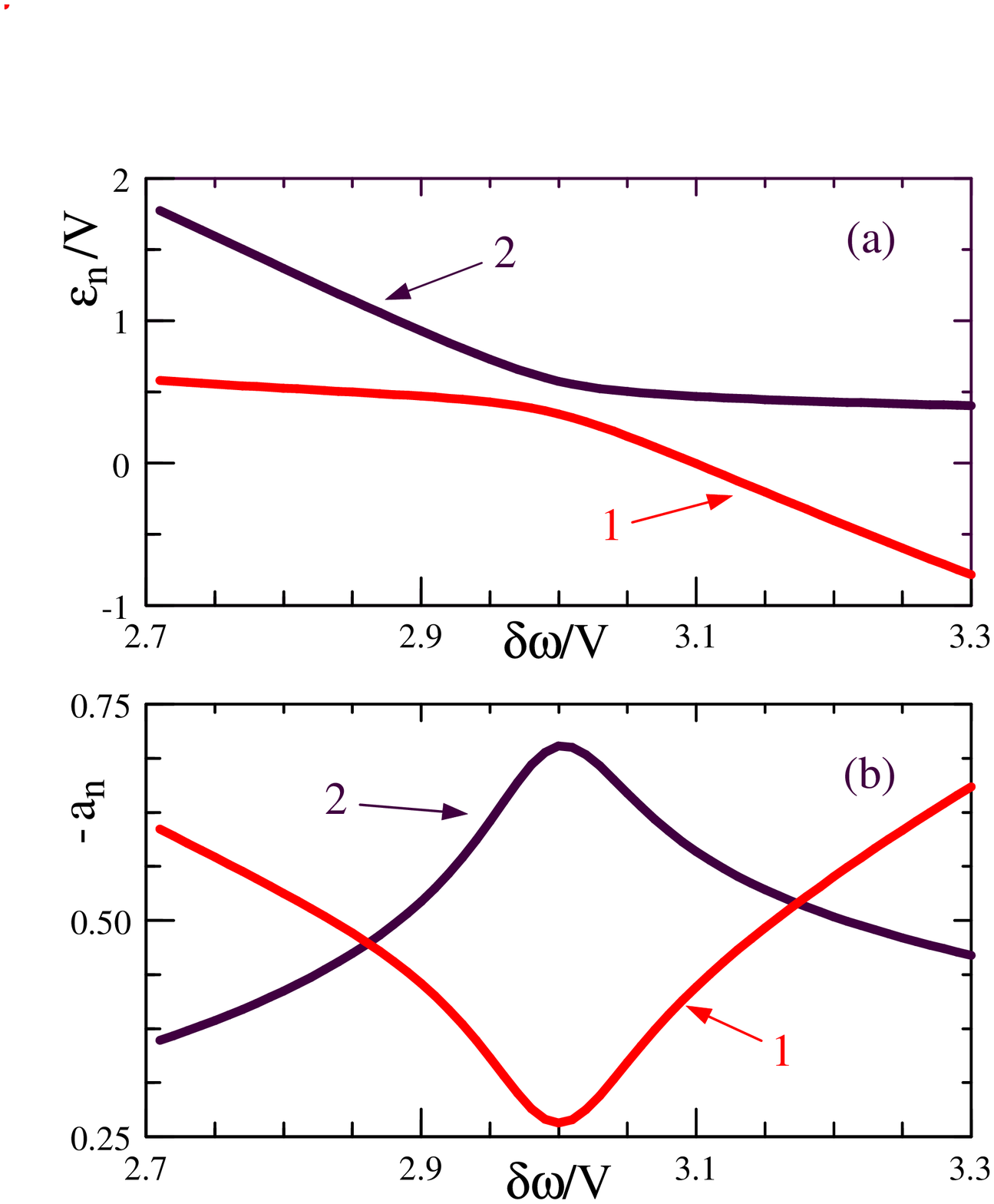}
\caption{(a) Anticrossing of the quasienergy levels $\varepsilon_0$
and $\varepsilon_5$ with varying frequency $\delta\omega$ for the
field amplitude $f/f_{5}=0.75$, where the scaling field $f_5$ is given
by Eq.~(\protect\ref{Rabi}) with $N=5$.  In the limit $f\to 0$ the
levels cross for $\delta\omega=\delta\omega_5=3V$; (b) The reduced
vibration amplitudes $a_0$ and $a_5$ for the same adiabatic states.
For $(\delta\omega_5-\delta\omega)/\Omega_R\gg 1$ the states described
by lines 1 and 2 are close to the Fock states of the oscillator
$|0\rangle$ and $|5\rangle$, respectively.}
\label{fig:anticrossing}
\end{figure}

Multiphoton mixing leads to splitting of the quasienergy levels
and the vibration amplitudes. It can be calculated by diagonalizing
the Hamiltonian (\ref{Hamiltonian_working}) and is shown in
Fig.~\ref{fig:anticrossing} as a function of frequency detuning
$\delta\omega$. One of the involved resonating states is the ground state of
the oscillator $n=0$ in the limit $f\to 0$.

The minimal splitting of the levels $\varepsilon_0$ and
$\varepsilon_N$ is given by the multiphoton Rabi frequency
$\Omega_R$. For weak field it can be obtained from
Eq.~(\ref{Hamiltonian_working}) by perturbation theory
\cite{Bloembergen76}. To the lowest order in $f/\delta\omega_N$
\begin{equation}
\label{Rabi_exact}
\Omega_R= 2f\,|2f/V|^{N-1}N^2\left(N!\right)^{-3/2}.
\end{equation}
For $N \gg 1$ this expression becomes
\begin{eqnarray}
\label{Rabi}
&&\Omega_R=V\,(f/f_N)^N N^{5/4}(2\pi)^{-3/4},\nonumber\\
&&f_N=|V|\,N^{3/2}\exp(-3/2)/2.
\end{eqnarray}

\begin{figure}
\includegraphics[width=2.4in]{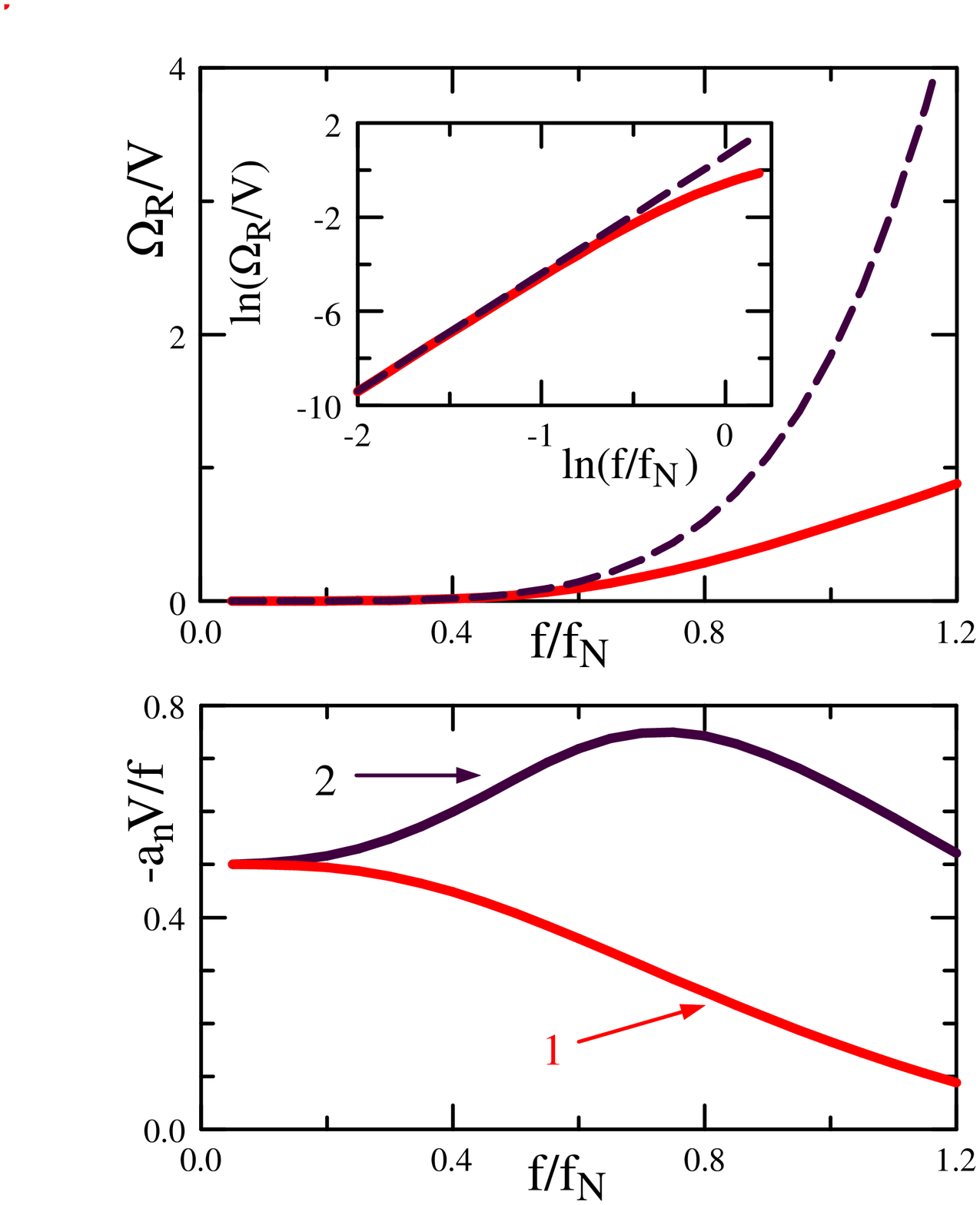}
\caption{{\it Upper panel:} field-induced splitting of the quasienergy
levels $n=0$ and $n = N=5$ for resonant driving frequency,
$\delta\omega=\delta\omega_5=3V$. The splitting gives the $5$-photon
Rabi frequency $\Omega_R$. The dashed line shows the weak-field
perturbation theory (\protect\ref{Rabi_exact}). {\it Lower panel:}
splitting of the reduced amplitudes of forced vibrations in the
corresponding Floquet states. The curve labelling coincides with that
in Fig.~\protect\ref{fig:anticrossing}.} \label{fig:rabi}
\end{figure}

The Rabi frequency depends on $N$ exponentially, $\Omega_R\propto
f^N$. In the case $N=5$ it is shown in
Fig.~\ref{fig:rabi} . One can see from this figure that
Eq.~(\ref{Rabi_exact}) works well in the whole range of the field
amplitudes $f/f_N\lesssim 0.5$. 
For larger fields $\Omega_R$ depends on $f$ much weaker than the
asymptotic expression (\ref{Rabi}) \cite{Dyakonov86}.

The most interesting feature of Fig.~\ref{fig:anticrossing} is the
antiresonant splitting of the amplitudes. It occurs at the adiabatic
passage of $\delta\omega$ through resonance, where the system switches
between the ground and excited states. In particular, the amplitude
displays an antiresonant dip if the oscillator is mostly in the ground
state for $(\delta\omega-\delta\omega_N)/V <1$ or in the state $N$ for
$(\delta\omega-\delta\omega_N)/V >1$. The magnitude and sharpness of
the dip are determined by $\Omega_R/V$ and depend very strongly on the
field and $N$. With decreasing $\Omega_R/V$ the dip (and peak) start
looking like cusps located at resonant frequency. The amplitude
splitting as function of the field is shown in Fig.~\ref{fig:rabi}.

To explain this behavior we note that, for $\delta\omega
=\delta\omega_N$, the field leads to two major effects. One is
resonant mixing of the wave functions into symmetric and antisymmetric
combinations $|n,N-n\rangle_{\pm}=\left(|n\rangle_0 \pm
|N-n\rangle_0\right)/\sqrt 2$ with quasienergies $\varepsilon_{n\pm}$
($|\cdot\rangle_0$ are eigenfunctions of the operator $\hat
n=a^{\dagger}a$).  The second effect is nonresonant mixing of the
states $|n,N-n\rangle_{\pm}$ with different $n$, which leads to
nonzero expectation values of the vibration amplitudes.

To first order in $f$, the vibration amplitudes $a_{0\pm}=\,
_{\pm}\!\langle 0,N|a|0,N\rangle_{\pm}$ are determined by nonresonant
mixing of the states $|0,N\rangle_{\pm}$ with $|1,N-1\rangle_{\pm}$
and $|N+1\rangle_0$. For comparatively weak fields, the level
splitting $\varepsilon_{1+}-\varepsilon_{1-} \propto
\Omega_R(\delta\omega/f)^2$ largely exceeds the splitting
$\varepsilon_{0+}-\varepsilon_{0-}= \Omega_R$. Then from perturbation
theory $a_{0+}-a_{0-} \propto f
[(\varepsilon_0-\varepsilon_{1+})^{-1}-(\varepsilon_0-\varepsilon_{1-})^{-1}]
\propto (f/V)^{N-1}$. This scaling describes the resonant small-field
amplitude splitting in Fig.~\ref{fig:rabi} extremely well (the
prefactor is determined by the admixture of states
$|n,N-n\rangle_{\pm}$ with $n>1$ and will be discussed elsewhere).

The simultaneous degeneracy of quasienergies and vibration amplitudes for
many pairs of states in a broad field range can be shown analytically
in the case where the oscillator dynamics is described by the WKB
approximation. This approximation applies for
\begin{equation}
\label{lambda_parameter}
\lambda \ll 1,\qquad\lambda =
V/(2\,\delta\omega).
\end{equation}

It is convenient to introduce the reduced coordinate and momentum of
the oscillator in the rotating frame
%
\[Q=(V/4\delta\omega)^{1/2}(a+a^{\dagger}),\quad
P=-i(V/4\delta\omega)^{1/2}(a-a^{\dagger})\]
%
with the commutator $[P,Q]=-i\lambda$. In these variables, in the
neglect of terms $\propto \lambda$, the Hamiltonian
(\ref{Hamiltonian_working}) becomes
$H_0=2(\delta\omega)^2V^{-1}[g(Q,P)-1/4]$, where
\begin{equation}
\label{g}
g(Q,P)=(Q^2+P^2-1)^2/4-\beta^{1/2}Q.
\end{equation}
Here $\beta=f^2V/(\delta\omega)^3$ is the reduced field intensity.

\begin{figure}[h]
\includegraphics[width=2.5in]{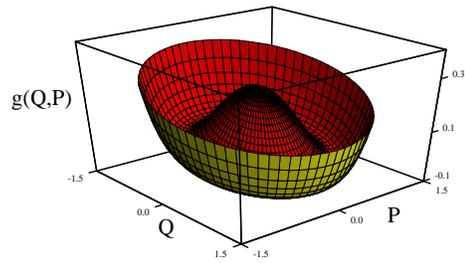}
\caption{The classical quasienergy of the oscillator
(\protect\ref{g}). The plot refers to the reduced field $\beta =
2/27$.} \label{fig:g_function}
\end{figure}

The function $g(Q,P)$ is illustrated in Fig.~\ref{fig:g_function};
$\delta\omega\, g(Q,P)$ is the classical Hamiltonian in the RWA, it
gives the quasienergy of the oscillator \cite{Dyakonov86,DK79}; $Q,P$
are the canonical variables.  The minimum and local maximum of
$g(Q,P)$ correspond to the stable states of forced vibrations. They
coexist for $0<\beta<4/27$. For such $\beta$, in a certain range of
$g$ there are two Hamiltonian trajectories with the same $g$,
one on the internal ``dome'' and the other on the external part of the
surface $g(Q,P)$. We call them, respectively, internal and external
trajectories.

The external trajectory for given $g(Q,P)=g$ has the form
$Q(t)=\beta^{-1/2}[X^2(t)-g]$, with
\begin{equation}
\label{trajectory_explicit}
X(t)= \frac{c_2(c_1-c_3)-c_3(c_1-c_2){\rm sn}^2u}
{c_1-c_3-(c_1-c_2){\rm sn}^2u}.
\end{equation}
Here, ${\rm sn}\,u$ is the Jacobi elliptic function; the elliptic
modulus is $m=(c_1-c_2)(c_3-c_4)/(c_1-c_3)(c_2-c_4)$, and
$u=[(c_1-c_3)(c_2-c_4)]^{1/2}\delta\omega\,t/2$ is the appropriately
scaled time. The coefficients $c_1>c_2>c_3>c_4$ are the roots of the
polynomial $\beta(1+2x)-(x^2-g)^2$ ($X=c_1,c_2$ and
$X=c_3,c_4$ gives the turning points
$\dot Q=0$ on the external and internal trajectories,  respectively).

The internal trajectory $Q(t)$ is given by
Eq.~(\ref{trajectory_explicit}) with $u\to u+K+iK^{\prime}$, where
$K\equiv K(m)$ is the complete elliptic integral, and
$K^{\prime}\equiv K(1-m)$.

An immediate consequence of the analytical interrelation between the
external and internal trajectories is that the periods of motion along
them coincide \cite{DS88}. The vibration frequency is
$\omega(g)=\pi[(c_1-c_3)(c_2-c_4)]^{1/2}\delta\omega/2K$. When motion
is quantized, $\omega(g)$ gives the distance between the energy
levels. Therefore if, for some $\delta\omega$ and $\beta$, two levels
that correspond to the external and internal trajectories coincide
with each other, many levels will coincide pairwise as well. Resonant
multiphoton splitting of the levels is due to tunneling between the
external and internal parts of the surface $g(Q,P)$.

In the WKB approximation the expectation value $a_n$ in a quantum
state $|n\rangle$ is given by the period-averaged value $\langle
Q\rangle_g$ of the coordinate $Q$ on the appropriate classical
trajectory. The values of $\langle Q\rangle_g$ turn out to be the same
on the internal and external trajectories with the same $g$. Indeed,
their difference can be written as a contour integral of $X^2$ in the
$u$-plane. The contour is a parallelogram with vertices at $u=0,2K,
3K+iK^{\prime}$, and $K+iK^{\prime}$. The function $X^2$ has one
second order pole in this parallelogram. The residue is $\propto
c_1+c_2+c_3+c_4=0$. This proves that the vibration amplitudes of the
states with equal quasienergies are equal, in the neglect of
tunneling.

In order to observe the coherent multiphoton quantum effects, the Rabi
frequency $\Omega_R$ should exceed the decoherence rate. In the RWA,
relaxation of an oscillator can often be described by the master
equation for the density matrix $\rho$. In many cases of interest the
major source of decoherence is noise or a quantum field that modulates
the oscillator energy. Then the coupling operator is $\propto \hat
n$. If the modulation is fast compared to $\delta\omega, V$, the
master equation takes the form
\begin{equation}
\label{QKE}
\dot\rho = i[\rho,H_0]-\hat{\Gamma}_{\varphi}\rho,\qquad
\hat{\Gamma}\rho = \Gamma_{\varphi}\left[\hat n,\left[\hat
n,\rho\right]\right],
\end{equation}
where $\Gamma_{\varphi}$ characterizes the noise intensity. It gives
the phase diffusion coefficient of the oscillator.

From Eq.~(\ref{QKE}), for resonant $\delta\omega=\delta\omega_N$ the
major effect of phase diffusion is decay of the Rabi oscillations. For
$\Omega_R/\delta\omega_N\ll 1$ the decay rate is
$\Gamma_{\varphi}N^2$: this is simply the diffusion coefficient of the
phase difference of the Fock states of the oscillator $|0\rangle$ and
$|N\rangle$. More formally, this is the decay rate of the difference
$\rho_{++}-\rho_{--}$, where $\rho_{\nu\nu^{\prime}} =\,_\nu\!\langle
0,N|\rho|0,N\rangle_{\nu^{\prime}}$ ($\nu,\nu^{\prime}=\pm$).

The decoherence rate due to phase diffusion quickly increases with
$N$.  Our results demonstrate that the strong resonant amplitude
splitting occurs already for $N=5$. We emphasize that this is a
coherent quantum effect. It is qualitatively different from the
nonmonotonic field dependence of the stationary amplitude of a driven
damped oscillator for nonzero temperatures \cite{Vogel88,Peano04}.

The antiresonances in the vibration amplitude can be directly observed
in Josephson junctions (JJ's), which are well described by the model
of a nonlinear oscillator (\ref{Hamiltonian_working}). The amplitude
of forced vibrations of a JJ was recently measured and the vibration
bistability was demonstrated \cite{Siddiqi_ampl03}. The eigenfrequency
and nonlinearity of a JJ are controlled by the dc bias current $I_{\rm
dc}$. When $I_{\rm dc}$ is close to the critical current $I_0$, i.e.,
$\eta=(I_0-I_{\rm dc})/I_0\ll 1$, in Eq.~(\ref{Hamiltonian_working})
$\omega_0=\omega_p(2\eta)^{1/4}$ and $V=-5\hbar\omega_p^2/48E_J\eta$,
where $\omega_p$ is the JJ plasma frequency and $E_J=\hbar
I_0/2e$. The scaling RF current $I_N$ for a resonant $N$-photon
transition, which corresponds to the scaling field $f_N$ in
Eq.~(\ref{Rabi}), is

\[I_N=(5/48)I_c\exp(-3/2)(2N\hbar\omega_p/E_J)^{3/2}(2\eta)^{-7/8}.\]

For the RF current $I_{\rm RF}\sim I_N$ at resonant frequency
$\omega_F=\omega_0+V(N+1)/2$, the splitting of the vibration amplitude
is strong, see Figs.~\ref{fig:anticrossing},
\ref{fig:rabi}. Roughly, $I_N$ is inversely proportional to
$I_0-I_{\rm dc}$, as is also the distance $V/2$ between the resonant
values $\omega_F$.

Multiphoton Rabi oscillations in multilevel JJs can be also studied by
measuring the rate of tunneling decay induced by resonant RF pulses of
different length $t_{\rm RF}$. The effective potential of a JJ is
metastable, $\omega_0^2q^2/2-\alpha q^3/3$. The probability
of tunneling escape from an excited state is much larger than from the
ground state, and the higher the energy level the stronger is the
difference. Therefore at multiphoton resonance the escape rate should
oscillate as $\sin^2(\Omega_Rt_{\rm RF}/2)$. This approach has been used
to detect single-photon Rabi oscillations in strongly nonlinear
JJ's that have a small number of metastable states \cite{HanMartMoya}.

We note that the quantum effects that we discuss can be studied also
in nanomechanical systems where classical bistability of forced
vibrations was  observed \cite{Cleland04}.

In this paper we have shown that multiphoton response of a quantum
oscillator may display antiresonant dips (peaks) as a function of
frequency.  The effect is due to level anticrossing and coherent
mixing of many oscillator states at a time.  The shape and magnitude
of the dips (peaks) strongly depend on the field. We find the
multiphoton Rabi frequency and discuss the possibility to observe the
antiresonant response and multiphoton Rabi oscillations in Josephson
junctions and nanomechanical resonators.

The work of M.D. was supported in part by the NSF through grant
No. ITR-0085922. M. V. F. thanks the hospitality of Michigan State
University where this work was started, and the financial support
of SFB 691.

\end{document}